\documentclass[useAMS,usenatbib,a4paper]{mn2e}

\usepackage{amssymb}
\usepackage{amsmath}
\usepackage{graphicx}
\usepackage{subfigure}
\usepackage{txfonts}
\usepackage{ifpdf}
\ifpdf
  \usepackage[pdftex]{hyperref}
\else
  \usepackage[ps2pdf,colorlinks=true]{hyperref}
\fi

\newif\ifbw
\bwfalse

\newcommand{\fig}[1]{Fig.~#1}
\newcommand{\Fig}[1]{Fig.~#1}
\newcommand{\sectn}[1]{Sec.~#1}

\newcommand{\etal}{\mbox{\it et al.}}

\newcommand{\ie}{\mbox{\it i.e.}}

\newcommand{\cmb}{{CMB}}
\newcommand{\cmbtext}{{cosmic microwave background}}

\newcommand{\wmap}{{WMAP}}
\newcommand{\wmaptext}{{Wilkinson Microwave Anisotropy Probe}}

\newcommand{\lcdm}{\ensuremath{\Lambda}{CDM}}
\newcommand{\lcdmtext}{\ensuremath{\Lambda} Cold Dark Matter}

\newcommand{\morlet}{real Morlet}

\newcommand{\healpix}{{\tt HEALPix}}

\newcommand{\lambdaarch}{{LAMBDA}}
\newcommand{\lambdaarchtext}{{Legacy Archive for Microwave Background Data Analysis}}

\title[Non-Gaussianity in the \wmap\ 5-year data]
  {A high-significance detection of non-Gaussianity in the \wmap\ 5-year %
   data using directional spherical wavelets}

\author[J.~D.~McEwen \etal]
  {J.~D.~McEwen,$^1$\thanks{E-mail: mcewen@mrao.cam.ac.uk} 
   M.~P.~Hobson,$^1$ A.~N.~Lasenby$^1$ and D.~J.~Mortlock$^2$\\
  $^1$Astrophysics Group, 
      Cavendish Laboratory, J.~J.~Thomson Avenue,
      Cambridge CB3 0HE, UK\\
  $^2$Blackett Laboratory, Imperial College of Science, Technology and Medicine,
    Prince Consort Road, London SW7 2BW, UK}
\date{Accepted 29 April 2008. Received 29 April 2008; in original form
  14 March 2008}
\pagerange{\pageref{sec:intro}--\pageref{lastpage}} 
\pubyear{2008}

\def\LaTeX{L\kern-.36em\raise.3ex\hbox{a}\kern-.15em
    T\kern-.1667em\lower.7ex\hbox{E}\kern-.125emX}

\begin{document}
\maketitle

\begin{abstract}
  We repeat the directional spherical \morlet\ wavelet analysis, used
  to detect non-Gaussianity in the \wmaptext\ (\wmap) 1-year and
  3-year data \citep{mcewen:2005:ng,mcewen:2006:ng}, on the \wmap\
  5-year data.  The non-Gaussian signal {detect\-ed} previously is
  present in the 5-year data at a slightly increased statistical
  significance of approximately 99\%.  Localised regions that
  contribute most strongly to the non-Gaussian signal are found to be
  very similar to those detected in the previous releases of the
  \wmap\ data.  When the localised regions detected in the 5-year data
  are excluded from the analysis the non-Gaussian signal is eliminated.
\end{abstract}


\section{Introduction}
\label{sec:intro}

The statistics of the primordial fluctuations provide a useful
mechanism for distinguishing between various scenarios of the early
Universe, such as various models of inflation.  Furthermore, the
primordial fluctuations give rise to the anisotropies of the \cmbtext\
(\cmb), which may be observed directly.  In the simplest inflationary
scenarios, primordial perturbations seed Gaussian temperature
fluctuations in the \cmb\ that are statistically isotropic over the
sky.  However, this is not the case for non-standard inflationary
models or alternative models to inflation.  Evidence of primordial
non-Gaussianity in the \cmb\ temperature anisotropies would therefore
have profound implications for the standard cosmological model.

Initial analyses of the \wmaptext\ (\wmap) 1-year
\citep{bennett:2003a}, three-year \citep{hinshaw:2006} and five-year
\citep{hinshaw:2008} observations of the \cmb\ (hereafter referred to
as \wmap1, \wmap3 and \wmap5), performed by \citet{komatsu:2003},
\citet{spergel:2006} and \citet{komatsu:2008} respectively, find no
evidence for deviations from Gaussianity.  However, no one statistic
is sensitive to all possible forms of non-Gaussianity that may exist
in the \wmap\ data due to either foreground contamination, systematics
or of primordial origin.  It is therefore important to test the data
for deviations from Gaussianity using a range of different methods
and, indeed, many additional studies have been performed on the \wmap1
and \wmap3 data:
\citealt{bernui:2007,
cabella:2005,cayon:2005,chen:2005,
chiang:2003,chiang:2004,chiang:2006,coles:2004,cg:2003,
creminelli:2007,
cruz:2005,cruz:2006a,cruz:2006b,
dineen:2005,eriksen:2004,eriksen:2005,eriksen:2007,
gw:2003,gott:2007,hansen:2004,
hikage:2008,jeong:2007,
larson:2004,larson:2005,lm:2004,lew:2008,
mcewen:2005:ng,mcewen:2006:ng,mcewen:2006:bianchi,
mm:2004,medeiros:2006,
monteserin:2007,mw:2004,naselsky:2007,
raeth:2007,sadegh:2006,tojeiro:2005,
vielva:2003,wiaux:2006,wiaux:2008,yadav:2007}.
Deviations from Gaussianity have been detected in many of these works.
Although the \wmap5 data are consistent with previous releases, the
modelling of beams is improved considerably, new masks are defined and
a further two-years of observations mean that the \wmap5 data can
provide reliable confirmation of previous non-Gaussianity analyses.

In this article we focus on the detection of non-Gaussianity that we
made previously in the \wmap1 and \wmap3 data
\citep{mcewen:2005:ng,mcewen:2006:ng}.  The Kp0
mask provided for previous \wmap\ releases and used in our Gaussianity
analyses was constructed from the K-band \wmap\ observations, which
contain \cmb\ and foreground emission.  Consequently, the application
of this mask may introduce negative skewness in the distribution of
the \cmb\ \citep{komatsu:2008}.  Since our previous detections of
non-Gaussianity were observations of negative skewness in wavelet
coefficients computed from \wmap\ data masked in this manner, it is
prudent to readdress our analysis in light of the new \wmap\ data and
masks.  The remainder of this letter is organised as follows.  In
\sectn{\ref{sec:analysis}} we discuss the \wmap5 map considered
and present the results of the non-Gaussianity analysis.  Concluding
remarks are made in \sectn{\ref{sec:conclusions}}.

\section{Non-Gaussianity analysis and results}
\label{sec:analysis}

We repeat our non-Gaussianity analysis performed previously on the
\wmap1 and \wmap3 data \citep{mcewen:2005:ng,mcewen:2006:ng}, focusing
on the most significant detection of non-Gaussianity made in the
skewness of real Morlet wavelet coefficients.  A detailed description
of the analysis procedure is presented in \citet{mcewen:2005:ng} and a
brief overview is also given in \citet{mcewen:2006:ng}.  Consequently,
we do not review the method in detail here but merely comment that it
involves a Monte Carlo analysis of real Morlet wavelet coefficients of
the data.  Twelve scales $a_i$ spaced equally between $50\arcmin$ and
$600\arcmin$ are considered.  Furthermore, the real Morlet wavelet
analysis probes directional structure in the data and we examine five
wavelet azimuthal orientations spaced equally in the domain $[0,\pi)$.  The
directional analysis is facilitated by our fast directional continuous
spherical wavelet transform code \citep{mcewen:2006:fcswt}, which is
based on the fast spherical convolution developed by
\citet{wandelt:2001}.

We consider the signal-to-noise ratio enhanced co-added map
constructed from the \wmap5 data (see
\citet{komatsu:2003}, \citet{mcewen:2005:ng} for descriptions of the co-added
map construction procedure).  Each simulated map used in the Monte
Carlo simulations is constructed in an analogous manner to the
co-added map constructed from the data.  A Gaussian \cmb\ realisation
is simulated from the theoretical \lcdmtext\ (\lcdm) power spectrum fitted
by the \wmap\ team \citep{dunkley:2008}.  Measurements made by the
various receivers are then simulated by convolving with realistic
beams and adding anisotropic noise for each receiver, where the beams
and noise properties used correspond to \wmap5 observations.  The
simulated observations for each receiver are then combined to give a
co-added map.  In this analysis we use the new KQ75 and KQ85 masks,
rather than the Kp0 mask used in our previous analyses.  The
construction of these new masks is discussed by \citet{gold:2008}. The
KQ75 mask is the more conservative of the two new masks and is
recommended for Gaussianity analyses \citep{komatsu:2008}.
Nevertheless, we consider both masks since these are the changes in
the \wmap5 data that are most likely to affect the results of our
analysis.

\begin{figure}
\centering
\ifpdf
\subfigure[KQ75 mask]{
\includegraphics[width=75mm]{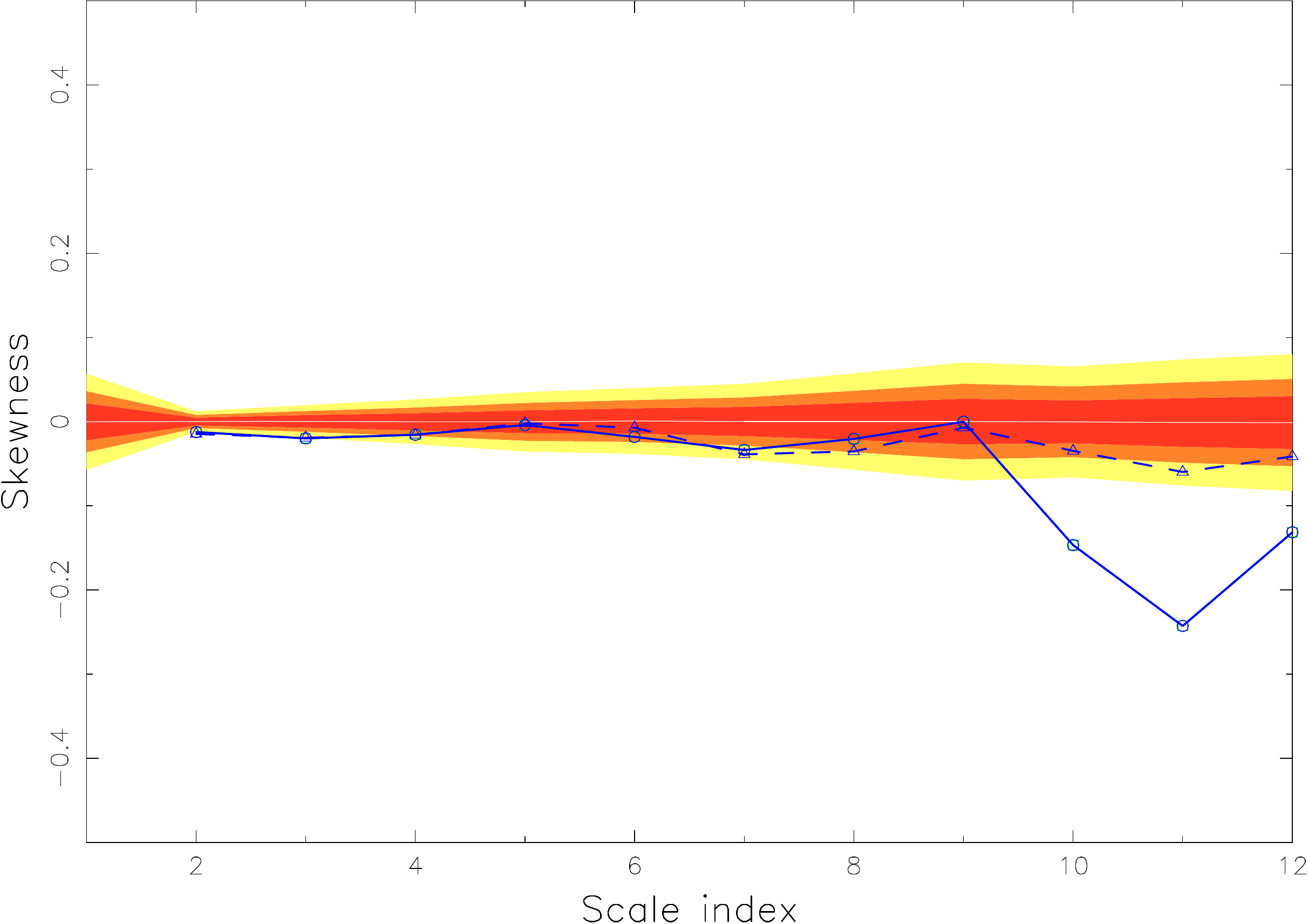}}
\subfigure[KQ85 mask]{
\includegraphics[width=75mm]{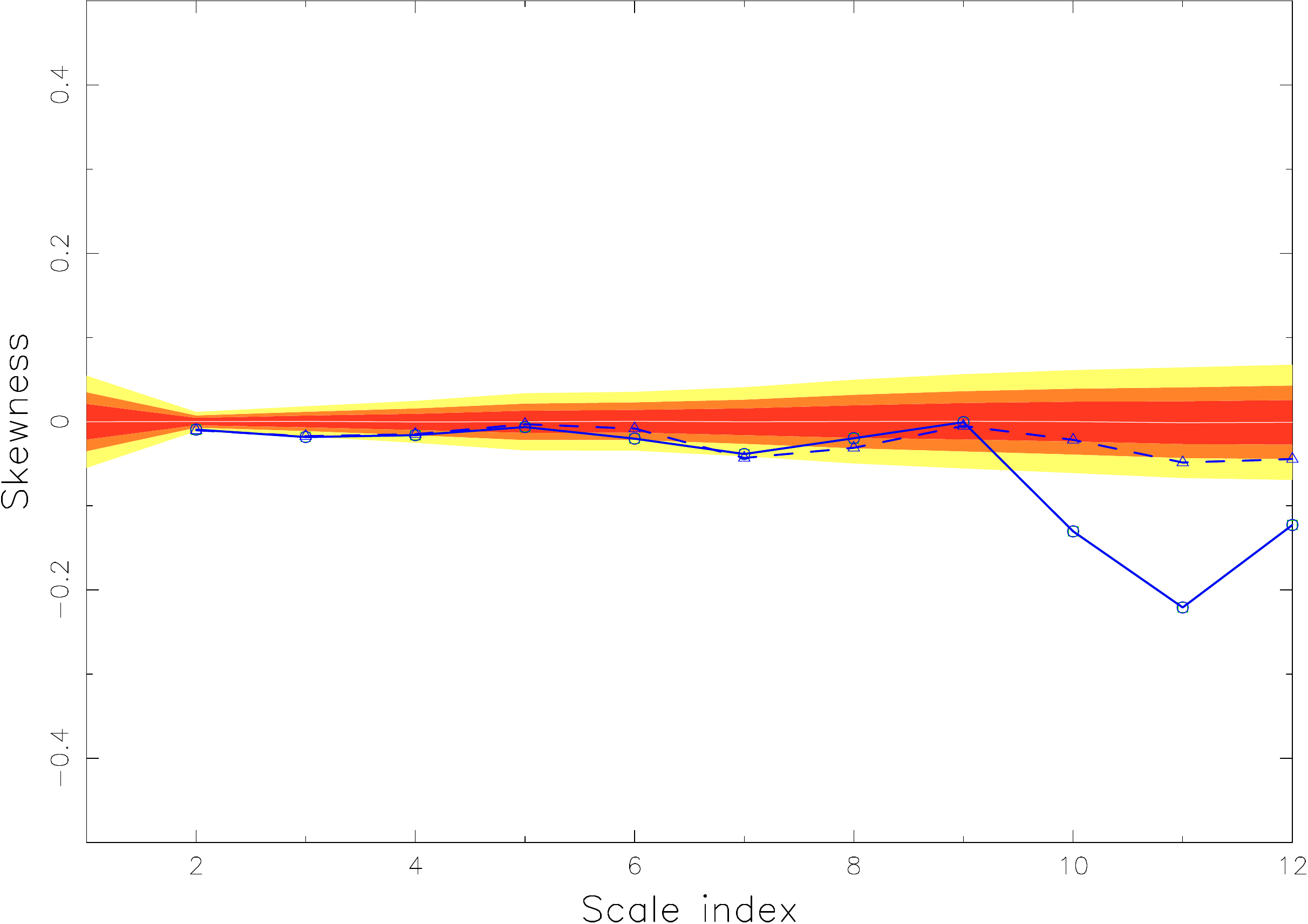}}
\else
\subfigure[KQ75 mask]{
\includegraphics[angle=-90,width=75mm]{figures/skewness_morlet_maskK75_dmask_ig02}}
\subfigure[KQ85 mask]{
\includegraphics[angle=-90,width=75mm]{figures/skewness_morlet_maskK85_dmask_ig02}}
\fi
\caption{Real Morlet wavelet coefficient skewness statistics
  ($\gamma=72^\circ$).  Points are plotted for the \wmap 5 data
  (solid, blue, circles), and the \wmap 5 data with localised regions
  removed (dashed, blue, triangles).  Confidence regions obtained from
  1000 \wmap 5 Monte Carlo simulations are shown for 68\% (red), 95\%
  (orange) and 99\% (yellow) levels, as is the mean (solid white
  line).  Panel~(a) shows the statistics computed using the KQ75 mask,
  whereas panel~(b) shows the statistics computing using the KQ85
  mask.}
\label{fig:stats}
\end{figure}

The skewness of the real Morlet wavelet coefficients of the co-added
\wmap5 map are displayed in \fig{\ref{fig:stats}}, with confidence
intervals constructed from 1000 Monte Carlo simulations consistent
with the \wmap5 observations also shown.  Only the plot corresponding
to the orientation of the maximum deviation from Gaussianity is shown.
The non-Gaussian signal present in previous releases of the \wmap\
data is clearly present in the \wmap 5 data for both choices of mask.
In particular, the large deviation on scale $a_{11}=550\arcmin$
and orientation $\gamma=72^\circ$ remains.

\begin{figure}
\centering
\ifpdf
\includegraphics[width=75mm]{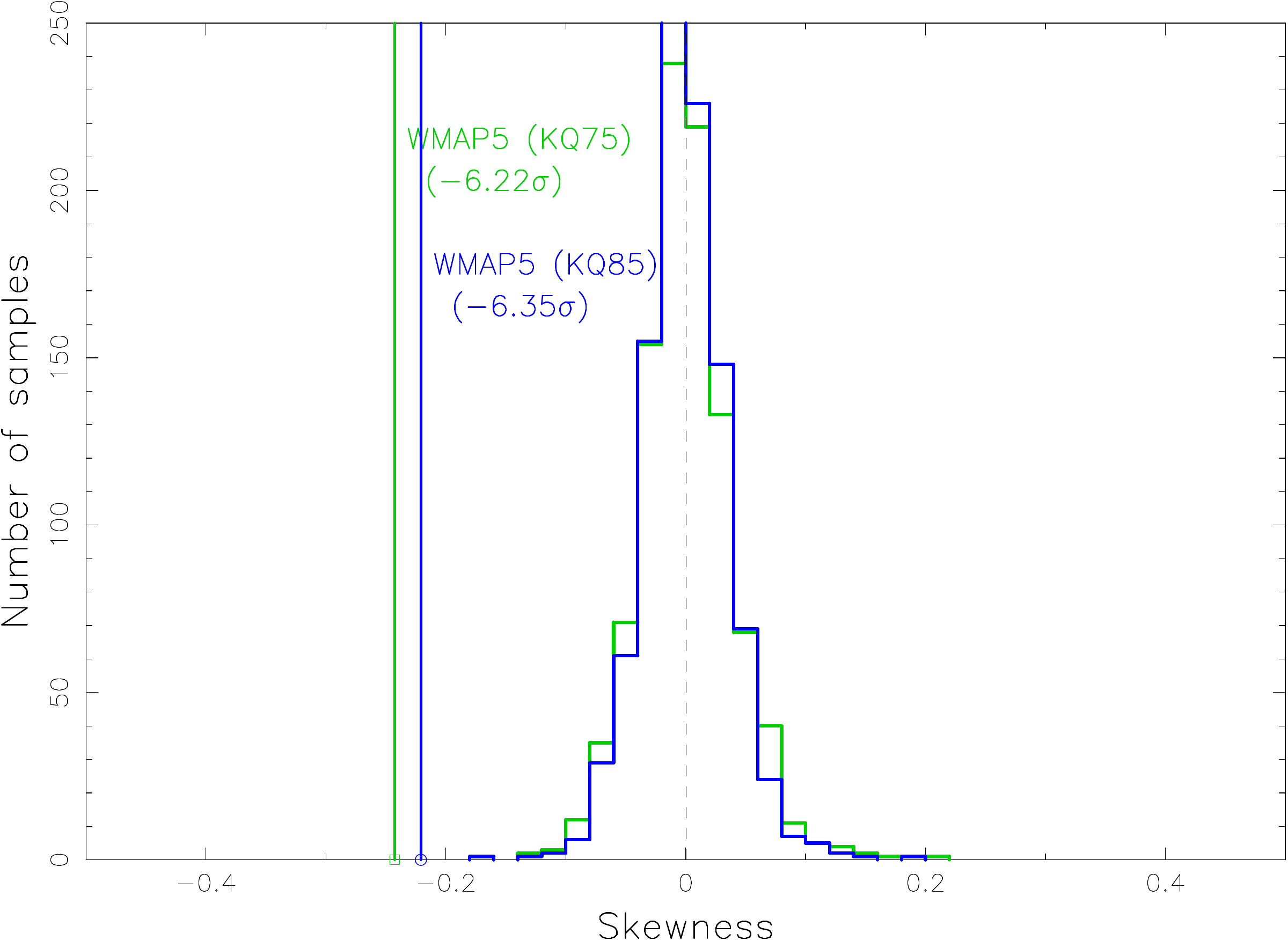}
\else
\includegraphics[angle=-90,width=75mm]{figures/hist2_skewness_morlet_ia11_ig02_maskboth}
\fi
\caption{Histograms of real Morlet wavelet coefficient skewness
  \mbox{($a_{11}=550\arcmin$; $\gamma=72^\circ$)} obtained from 1000
  \wmap 5 Monte Carlo simulations.  Histograms are plotted for
  statistics computed from the simulations using the KQ75 (green) and
  KQ85 (blue) masks.  The observed statistics for the \wmap 5 data
  with the KQ75 and KQ85 masks maps are shown by the green and blue
  lines respectively.  The number of standard deviations these
  observations deviate from the mean of the appropriate set of
  simulations is also displayed.}
\label{fig:hist}
\end{figure}

Next we consider in more detail the most significant deviation from
Gaussianity on scale $a_{11}=550\arcmin$ and orientation
$\gamma=72^\circ$.  \Fig{\ref{fig:hist}} shows histograms of this
particular statistic constructed from the \wmap5 Monte Carlo
simulations for both masks.  The skewness value measured from the data
is also shown on the plot, with the number of standard deviations each
observation deviates from the mean of the appropriate set of
simulations.  The distribution of this skewness statistic is not
significantly altered between simulations analyses with the KQ75 or
KQ85 masks.  The observed statistics for the data are also similar for
both masks.

\begin{figure}
\centering
\ifpdf
\includegraphics[width=75mm]{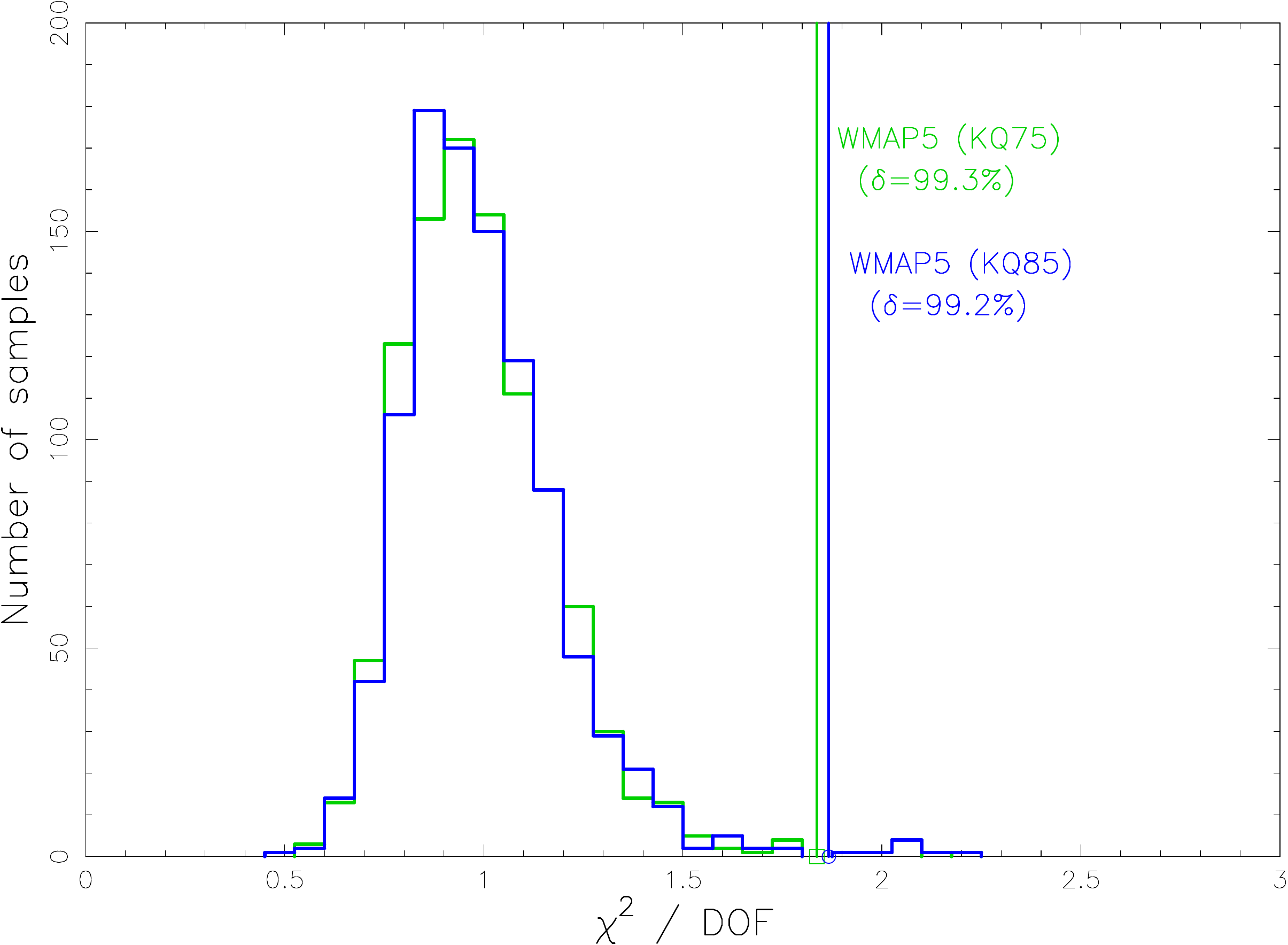}
\else
\includegraphics[angle=-90,width=75mm]{figures/histchi2_wmap5_maskboth_morlet_sTkT}
\fi
\caption{Histograms of normalised $\chi^2$ test statistics computed
  from real Morlet wavelet coefficient statistics obtained from 1000
  \wmap 5 Monte Carlo simulations.  Histograms are plotted for
  statistics computed from the simulations using the KQ75 (green) and
  KQ85 (blue) masks.  The $\chi^2$ value for the \wmap 5 data with the
  KQ75 and KQ85 masked maps are shown by the green and blue lines
  respectively.  The significance of these observations, computed from
  the appropriate set of simulations, is also displayed.}
\label{fig:chi2}
\end{figure}

To quantify the statistical significance of the detected deviation
from Gaussianity we consider two techniques.  The first technique
involves comparing the deviation of the skewness statistic computed
from the \wmap5 data on scale $a_{11}=550\arcmin$ and orientation
$\gamma=72^\circ$ to all statistics computed from the simulations.
This is a very conservative means of constructing significance levels.
The second technique involves performing a $\chi^2$ test.  The
$\chi^2$ value computed from the \wmap5 data is compared to $\chi^2$
statistics computed from the simulations.  In both of these tests we
relate the observation to all test statistics computed originally,
\ie\ to both skewness and kurtosis statistics.  For a more thorough
description of these techniques see \citet{mcewen:2005:ng}.
Using the first technique, the significance of the detection of
non-Gaussianity in the \wmap5 is made at $99.2\pm0.3$\% and
$99.1\pm0.3$\% using the KQ75 and KQ85 masks respectively.
The distribution of $\chi^2$ values obtained from the Monte Carlo
simulations is shown in \fig{\ref{fig:chi2}}.  The $\chi^2$ value
obtained for the data is also shown on the plot.  Again, the
distribution and value observed in the data is not altered
significantly when using the different masks.  Computing the
significance of the detection of non-Gaussianity directly from the
$\chi^2$ distributions and observations, the significance of the
detection in the \wmap5 data is made at $99.3\pm0.3$\% and
$99.2\pm0.3$\% using the KQ75 and KQ85 masks respectively.
Using both of the techniques outlined above the detection of
non-Gaussianity made in the \wmap5 data is made at a slightly higher
significance than in previous releases of the data.  Nevertheless, the
same non-Gaussian signal appears to be present.

\begin{figure}
\centering
\subfigure[KQ75 mask]{
\includegraphics[width=75mm]{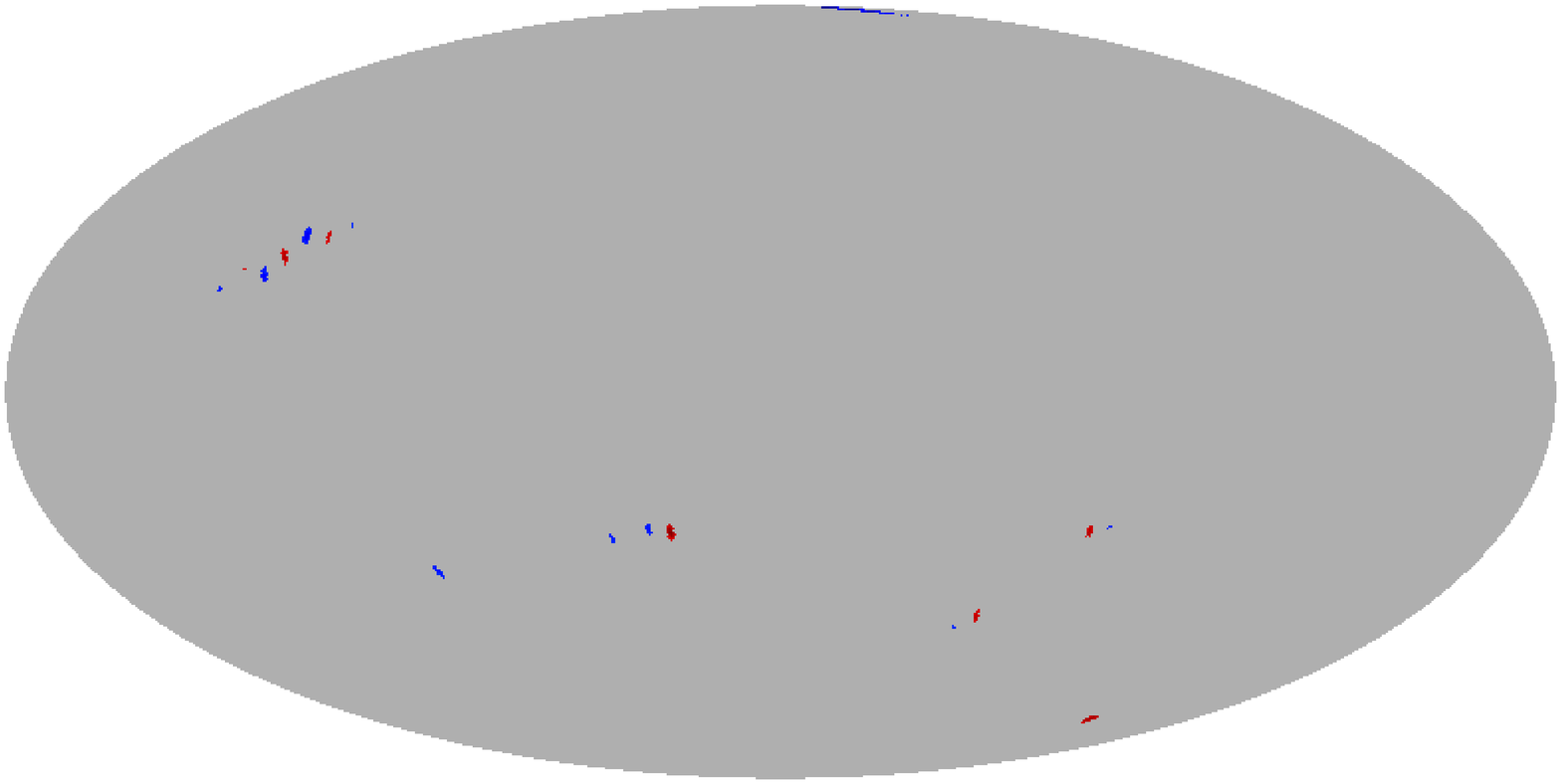}}
\subfigure[KQ85 mask]{
\includegraphics[width=75mm]{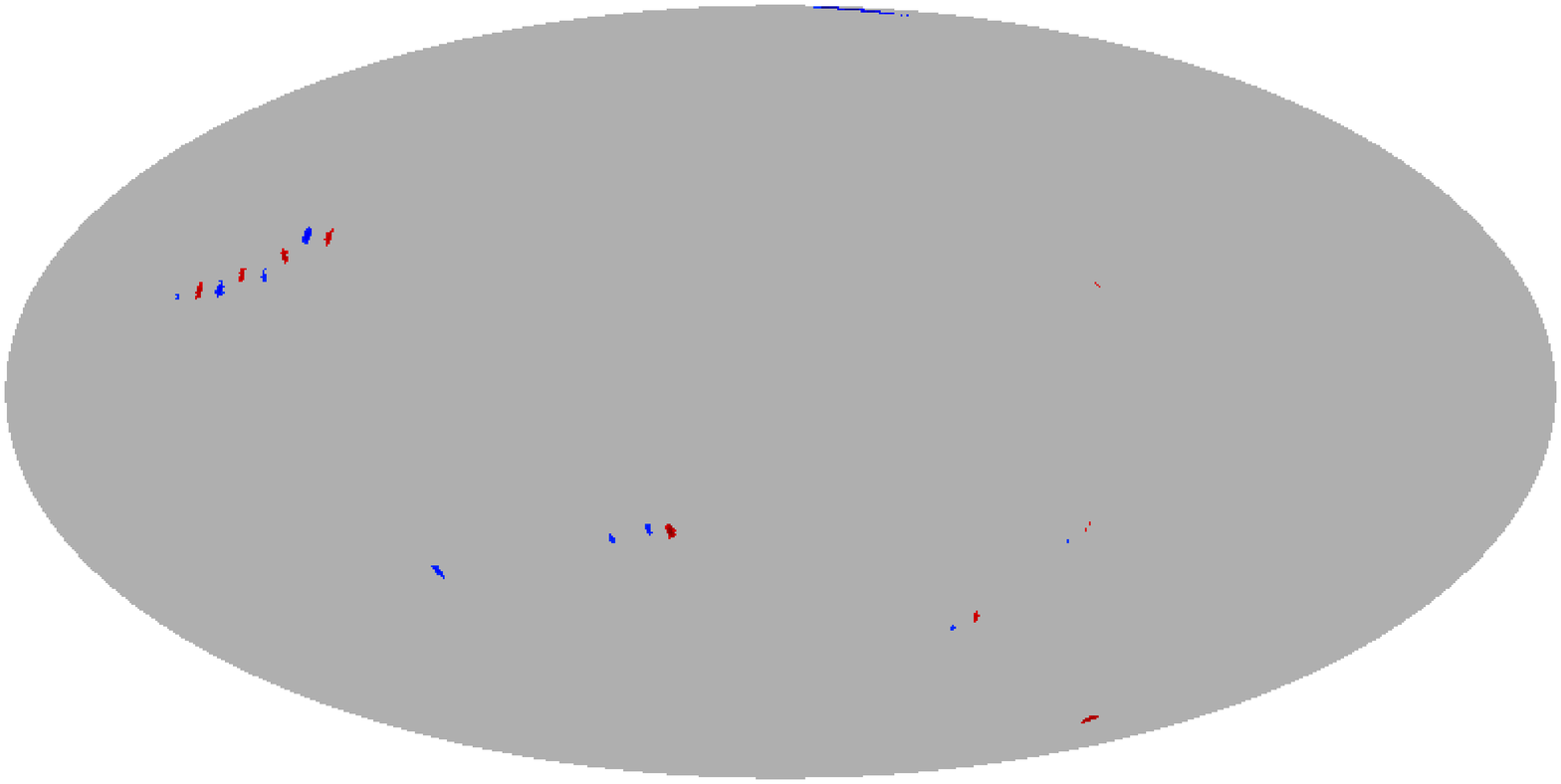}}
\caption{Real Morlet wavelet thresholded spherical wavelet coefficient maps
  \mbox{($a_{11}=550\arcmin$; $\gamma=72^\circ$)}.  To localise
  most likely deviations from Gaussianity on the sky, the original
  wavelet coefficient map is thresholded so that only those
  coefficients above $3\sigma$ (in absolute value) remain.  All sky
  maps are illustrated in Galactic coordinates, with the Galactic
  centre in the middle.}
\label{fig:coeff}
\end{figure}

The wavelet analysis allows one to localise those regions on the sky
that contribute most significantly to deviations from Gaussianity
\citep{mcewen:2005:ng}.  In \fig{\ref{fig:coeff}} we plot the
thresholded wavelet coefficients corresponding to the most significant
detection of non-Gaussianity made on scale $a_{11}=550\arcmin$ and
orientation $\gamma=72^\circ$.\footnote{We make corresponding
  localised region masks available publicly from
  \url{http://www.mrao.cam.ac.uk/~jdm57/} so that other researchers
  may determine whether these regions are responsible for detections
  of non-Gaussianity made with other analysis techniques.}  These
localised regions match the localised regions detected in the \wmap1
and \wmap3 data closely.  When excluding localised regions from the
initial analysis, the highly significant non-Gaussian signals present
previously are eliminated (see \fig{\ref{fig:stats}}).

In our previous non-Gaussianity analyses we concluded that noise was
not atypical in the localised regions detected \citep{mcewen:2005:ng}.
Moreover, we also concluded that foregrounds and systematics were not
the likely source of the detected non-Gaussianity
\citep{mcewen:2006:bianchi}.  The localised regions detected in the
\wmap5 data have not changed markedly to those detected previously and
foregrounds and systematics are treated more thoroughly, hence we do
not expect these findings to alter in the \wmap5 data.

\section{Conclusions}
\label{sec:conclusions}

In this work we have repeated our non-Gaussianity analysis on the
\wmap5 data.  The non-Gaussian signal detected previously remains
present in the \wmap5 data.  The possible introduction of negative
skewness in the \wmap\ data by the application of the Kp0 mask
\citep{komatsu:2008} appears not to be responsible for our
non-Gaussian signal.  Non-Gaussianity is detected at significance
levels of $99.2\pm0.3$\% and $99.1\pm0.3$\% using the KQ75 and KQ85
masks respectively, when using our conservative method for
constructing significance measures.  Using our second method, which is
based on a $\chi^2$ analysis, the significance of the detection is
made at $99.3\pm0.3$\% and $99.2\pm0.3$\% using the KQ75 and KQ85
masks respectively.  These detections of deviations from Gaussianity
are made at a slightly higher significance in the \wmap5 data than in
previous releases.  We have no intuitive explanation for this
marginal rise in significance.  The most likely sources of
non-Gaussianity that were localised on the sky in the \wmap5 data
match those regions detected from previous releases of the data
reasonable closely (and are made available publicly).

It is interesting to note that the highly significant detection of
primordial non-Gaussianity made with the bispectrum by
\citet{yadav:2007} is sensitive to skewness, which is also the type of
non-Gaussianity detected with our real Morlet wavelet analysis.  To
test whether these two analyses detect the same source of
non-Gaussianity, one could remove the localised regions that we detect
and repeat the analysis performed by \citet{yadav:2007} to see if
their detection of non-Gaussianity remains.  This analysis is
currently being performed by Yadav \& Wandelt (private communication).

\section*{Acknowledgements}

Some of the results in this paper have been derived using the
\healpix\footnote{\url{http://healpix.jpl.nasa.gov/}} package
\citep{gorski:2005}.  We acknowledge the use of the
\lambdaarchtext\footnote{\url{http://lambda.gsfc.nasa.gov/}}
(\lambdaarch).  Support for \lambdaarch\ is provided by the NASA
Office of Space Science.

\bibliographystyle{mymnras_eprint}
\bibliography{bib}

\label{lastpage}
\end{document}